\documentclass[10pt,twocolumn]{article}
\usepackage{times}

\usepackage[draft]{fixme}
\usepackage[latin1]{inputenc}
\usepackage{algorithm}
\usepackage[noend]{distribalgo}
\usepackage{amssymb}
\usepackage{amsthm}
\usepackage{multirow}

\theoremstyle{plain}

\theoremstyle{definition}
\newtheorem{definition}{Definition}

\usepackage{color} 

   \usepackage[pdftex]{graphicx}

\usepackage{algorithm}
\usepackage{algorithmic}
\usepackage{subfigure}
\usepackage[colorinlistoftodos]{todonotes}
\newcommand{\static}{early}
\newcommand{\Static}{Early}

\newcommand{\Dynamic}{Late}

\begin{document}

\title{\Static\ Scheduling in Parallel State Machine Replication}

\author{Eduardo Alchieri$^1$, Fernando Dotti$^2$ and Fernando Pedone$^3$ \\
\small {\em  $^1$Departamento de Ciência da Computação -- Universidade de Bras\'{i}lia, Brazil} \\
\small {\em $^2$Escola Polit\'{e}cnica -- Pontif\'{i}cia Universidade Cat\'{o}lica do Rio Grande do Sul, Brazil} \\
\small {\em $^3$Universit\'{a} della Svizzera italiana (USI), Switzerland} \\ [2mm]
}
\date{}
\maketitle

\begin{abstract}

State machine replication is standard approach to fault tolerance.
One of the key assumptions of state machine replication is that replicas must execute operations deterministically and thus serially.
To benefit from multi-core servers, some techniques allow concurrent execution of operations in state machine replication.
Invariably, these techniques exploit the fact that independent operations (those that do not share any common state or do not update shared state) can execute concurrently.
A promising category of solutions trades scheduling freedom for simplicity.
This paper generalizes this category of scheduling solutions.
In doing so, it proposes an automated mechanism to schedule operations on worker threads at replicas.
We integrate our contributions to a popular state machine replication framework and experimentally compare the resulting system to more classic approaches.

\end{abstract}

%
%
%
%

\section{Introduction}

This paper investigates a new class of scheduling protocols for high performance parallel state machine replication.
State machine replication (SMR) is a well-established approach to fault tolerance~\cite{Lam78,Sch90}.
The basic idea is quite simple: server replicas execute client requests deterministically and in the same order.
Consequently, replicas transition through the same sequence of states and produce the same sequence of outputs.
State machine replication can tolerate a configurable number of faulty replicas.
Moreover, application programmers can focus on the inherent complexity of the application, while remaining oblivious to the difficulty of handling replica failures~\cite{Fis85}.
Not surprisingly, the approach has been successfully used in many contexts (e.g., \cite{B06,scatter,CDE12}).

Modern multi-core servers, however, challenge the state machine replication model since deterministic execution of requests often translates into single-threaded replicas.
In search of solutions to this performance limitation, a number of techniques have been proposed to allow multi-threaded execution of requests at replicas (e.g., \cite{guo2014rex, kapitza2010storyboard, kapritsos2012all, Mendizabal17}).
Techniques that introduce concurrency in state machine replication build on the observation that \emph{independent} requests can execute concurrently while \emph{conflicting} requests must be serialized and executed in the same order by the replicas. 
Two requests conflict if they access common state and at least one of them updates the state, otherwise the requests are independent. 

An important aspect in the design of multi-threaded state machine replication is how to schedule requests for execution on worker threads.
Proposed solutions fall in two categories.
With \emph{late scheduling}, requests are scheduled for execution after they are ordered across replicas.
Besides the aforementioned requirement on conflicting requests, there are no further restrictions on scheduling.
With \emph{early scheduling}, part of the scheduling decisions are made before requests are ordered.
After requests are ordered, their scheduling must respect these restrictions.

Since late scheduling has fewer restrictions, it allows more concurrency than early scheduling.
Hence, a natural question is why would one resort to an early scheduling algorithm instead of a late scheduling algorithm?
The answer is that the cost of tracking dependencies among requests in late scheduling may outweigh its gains in concurrency.
In~\cite{Kot04}, for example, each replica has a directed dependency graph that stores not-yet-executed requests and the order in which conflicting requests must be executed.
A scheduler at the replica delivers requests in order and includes them in the dependency graph.
Worker threads remove requests from the graph and execute them respecting their dependencies.
Therefore, scheduler and workers contend for access to the shared graph.

By restricting concurrency, scheduling can be done more efficiently.
Consider a service based on the typical readers-and-writers concurrency model.
Scheduling becomes simpler if reads are scheduled on any one worker thread and writes are scheduled on all worker threads.
To execute a write, all worker threads must coordinate (e.g., using a barrier) so that no read is ongoing and only one worker executes the write.
This scheme is more restrictive than the one using the dependency graph because it does not allow concurrent writes, even if they are independent.
Previous research has shown that early scheduling can outperform late scheduling by a large margin, specially in workloads dominated by read requests \cite{Par14,Mendizabal17}.
These works assume a simple form of early scheduling that focuses on the readers-and-writers concurrency model.

This paper generalizes early scheduling in state machine replication.
We follow a three-step approach.
First, by using the notion of classes of requests we show how a programmer can express concurrency in an application.
In brief, the idea is to group service requests in classes and then specify how classes must be synchronized.
For example, we can model readers-and-writers with a class of read requests and a class of write requests.
The class of write requests conflicts with itself and with the class of read requests.
This ensures that a write is serialized with reads and with other writes.
We also consider more elaborate concurrency models that assume sharded application state with read and write operations within and across shards.
These models allow concurrent reads and writes.

Second, we show how one can automatically map request classes to worker threads. 
The crux of our technique is the formulation of the classes-to-threads mapping as an optimization problem.
Mapping classes to threads and deciding how to synchronize worker threads within and across classes is a non-trivial problem.
One interesting finding is that it may be advantageous to serialize an inherently concurrent class (i.e., with read requests only) to capitalize on more significant gains from increased concurrency in other classes, as we explain in detail in the paper.
Finally, our optimization formulation accounts for skews in the workload.

Third, we have fully implemented our techniques and conducted a number of experiments with different services and workloads, and compared the performance of our early scheduling technique to a late scheduling algorithm.

The paper continues as follows.
Section~\ref{sec:model} introduces the system model and consistency criteria. 
Section~\ref{sec:paralellismInSMR} provides some background on parallel approaches to state machine replication.
Section~\ref{sec:parallel} generalizes early scheduling.
Section~\ref{sec:optm} describes our optimization model.
Section~\ref{exp} reports on our experimental evaluation. 
Section~\ref{rw} surveys related work and Section~\ref{sec:conc} concludes the paper.

\section{System model and consistency}
\label{sec:model}

We assume a distributed system composed of interconnected processes that communicate by exchanging messages. 
There is an unbounded set of client processes and a bounded set of replica processes. 
The system is asynchronous: there is no bound on message delays and on relative process speeds. 
We assume the crash failure model and exclude arbitrary behavior. 
A process is \emph{correct} if it does not fail, or \emph{faulty} otherwise. 
There are up to $f$ faulty replicas, out of $2f+1$ replicas. 

Processes have access to an atomic broadcast communication abstraction, defined by primitives $broadcast(m)$ and $deliver(m)$, where $m$ is a message. 
Atomic broadcast ensures the following 
properties \cite{defago2004total,hadzilacos1993fault}\footnote{Atomic broadcast needs additional synchrony assumptions to be implemented~\cite{Cha96,Fis85}. These assumptions are not explicitly used by the protocols proposed in this paper.}:
\begin{itemize}
\item \emph{Validity}: If a correct process broadcasts a message $m$, then it eventually delivers $m$.
\item \emph{Uniform Agreement}: If a process delivers a message $m$, then all correct processes eventually deliver $m$.
\item \emph{Uniform Integrity}: For any message $m$, every process delivers $m$ at most once, and only if $m$ was previously broadcast by a process.
\item \emph{Uniform Total Order}: If both processes $p$ and $q$ deliver messages $m$ and $m'$, then $p$ delivers $m$ before $m'$, if and only if $q$ delivers 
$m$ before $m'$.
\end{itemize}

Our consistency criterion is \emph{linearizability}.
A linearizable execution satisfies the following requirements~\cite{Her90}:
\begin{itemize}
\item \label{lable:rtorder} It respects the real-time ordering of operations across all clients. There exists a real-time order among any two operations if one operation finishes at a client before the other operation starts at a client.
\item \label{lable:seqsem} It respects the semantics of the operations as defined in their sequential execution.
\end{itemize}


\section{Background}
\label{sec:paralellismInSMR}

Parallel SMR exploits the fact that strong consistency does not always require all operations to be executed in the same order at all replicas.
In this section, we formalize this idea and present two categories of SMR systems that exploit parallelism.

%

\subsection{The notion of conflict}
\label{sec:smrconflicts}

State machine replication determines how service operations must be propagated to and executed by the replicas.
Typically, (i) every correct replica must receive every operation;
(ii) no two replicas can disagree on the order of received and executed operations; and
(iii) operation execution must be deterministic: replicas must reach the same state and produce the same output upon executing the same sequence of operations.
Even though executing operations in the same order and serially at replicas is sufficient to ensure consistency, it is not always necessary, as we now explain.

Let $R$ be the set of requests available in a service (i.e., all the requests that a client can issue).
A request can be any deterministic computation involving objects that are part of the application state.
We denote the sets of application objects that replicas read and write when executing $r$ as $r$'s \emph{readset} and \emph{writeset}, or $RS(r)$ and $WS(r)$, respectively.
We define the notion of conflicting requests as follows.

\begin{definition}[Request conflict]
\label{def.rconflict}
The conflict relation $\#_R \subseteq R\times R$ among requests is defined as 
\[ 
(r_i,r_j) \in \#_R\ \textrm{iff} 
\left(
\begin{array}{c}
RS(r_i) \cap WS(r_j) \neq \emptyset \\
\lor \\
WS(r_i) \cap RS(r_j) \neq \emptyset \\
\lor \\
WS(r_i) \cap WS(r_j) \neq \emptyset
\end{array}
\right)
\]
\end{definition}

Requests $r_i$ and $r_j$ \emph{conflict} if $(r_i,r_j) \in \#_R$.
We refer to pairs of requests not in $\#_R$ as \emph{non-conflicting} or \emph{independent}.
Consequently, if two requests are independent (i.e., they do not share any objects or only read shared objects), then the requests can be executed concurrently at replicas (e.g., by different worker threads at each replica).
Concurrent execution of requests raises the issue of how requests are scheduled for execution on worker threads.
We distinguish between two categories of protocols.

\subsection{\Dynamic\ scheduling}

In this category of protocols, replicas deliver requests in total order and then a scheduler at each replica assigns requests to worker threads for execution.
The scheduler must respect dependencies between requests.
More precisely, if requests $r_i$ and $r_j$ conflict and $r_i$ is delivered before $r_j$, then $r_i$ must execute before $r_j$.
If $r_i$ and $r_j$ are independent, then there are no restrictions on how they should be scheduled (e.g., the scheduler can assign each request to a different worker thread).

CBASE~\cite{Kot04} is a protocol in this category.
In CBASE, a deterministic scheduler (or parallelizer) at each replica delivers requests in total order and includes them in a dependency graph.
In the dependency graph, vertices represent delivered but not yet executed requests and directed edges represent dependencies between requests.
Request $r_i$ depends on $r_j$ (i.e., $r_i \rightarrow r_j$ is an edge in the graph) if $r_i$ is delivered after $r_j$, and $r_i$ and $r_j$ conflict.

The dependency graph is shared with a pool of worker threads.
The worker threads choose requests for execution from the dependency graph respecting their interdependencies: a worker thread can execute a request if the request is not under execution and it does not depend on any requests in the graph.
After the worker thread executes the request, it removes the request from the graph and chooses another one.

%
%
%
%

\subsection{\Static\ scheduling}

With \static\ scheduling, a request is assigned to a worker thread (or to a pool of worker threads, from which one thread will execute the request) before it is ordered.
For example, one could establish that all requests that access object $x$ are executed by thread $t_0$ and all requests that access object $y$ are executed by thread $t_1$; requests that access both objects require threads $t_0$ and $t_1$ to coordinate so that only one thread executes the request (we detail this execution model in the next section).
The advantage of \static\ scheduling is that since scheduling decisions are simple (e.g., there is no dependency graph), the scheduler is less likely to become a performance bottleneck. 


Two variants of \static\ scheduling in state machine replication have been proposed.
In both cases, clients tag requests with the ids of the worker threads that will execute the requests.
In \cite{Alchieri2017}, a scheduler thread delivers requests in total order and assigns each request to the worker responsible for the execution of the request.
In \cite{Par14}, there is no scheduler involved; the atomic broadcast primitive uses the request tag to deliver requests directly to worker threads at the replicas.
Neither proposal explains how clients must tag requests to maximize concurrency.

\section{Parallelism with request classes}
\label{sec:parallel}

In this section, we introduce the notion of classes of requests and illustrate how they can be used to represent concurrency in three prototypical applications.

\subsection{Classes of requests}
\label{sec:classes}

\emph{Classes of requests} group requests and their interdependencies.
Each class has a descriptor and conflict information, as defined next.

\begin{definition}[Request classes]
\label{def.rClasses}
Recall that $R$ is the set of requests available in a service (c.f. \S\ref{sec:smrconflicts}).
Let $C = \{c_1, c_2, ..., c_{nc} \}$ be the set of class descriptors, where $nc$ is the number of classes.
We define request classes as $\mathcal{R} = C \rightarrow \mathcal{P}(C) \times \mathcal{P}(R)$,\footnote{We denote the power set of set $S$ as $\mathcal{P}(S)$.} 
that is, any class in $C$ may conflict with any subset of classes in $C$, and is associated to a subset of requests in $R$.
Moreover, we introduce the restriction that each request is associated to one and only one class.
\end{definition}

A request class set $\mathcal{R}$ is correct with respect to a request conflict $\#_R$ if every conflict in $\#_R$ is in $\mathcal{R}$:

\begin{definition}[Correct request classes]
\label{def.RCtoR}
Given a request conflict relation $\#_R$, a request class set $\mathcal{R}$ is correct with respect to $\#_R$ if for all $(r_i,r_j) \in \#_R$, with $i \not = j$, it follows that 
$\exists C_i \rightarrow (CC_i,CR_i) \in \mathcal{R}$ and $\exists C_j \rightarrow (CC_j,CR_j) \in \mathcal{R}$ such that:
(a)~$r_i \in CR_i$, 
(b)~$r_j \in CR_j$, 
(c)~$C_j \in CC_i$, and 
(d)~$C_i \in CC_j$.
\end{definition}

Intuitively, requests that belong to conflicting classes have to be serialized according to the total order induced by atomic broadcast.
Requests that belong to non-conflicting classes can be executed concurrently.
We differentiate between two types of conflicts involving classes.

\begin{itemize}
\item \emph{Internal conflicts:} We say that class $c$ has internal conflicts if $c$ contains requests that conflict. 
Requests in a class with internal conflict must be serialized and executed in the same order across replicas.  
If a class does not have internal conflicts, its requests can be processed concurrently.
\item \emph{External conflict:} If classes $c_i$ and $c_j$, $i\neq j$, conflict, then requests in $c_i$ and $c_j$ have to be serialized and executed in the same order by the replicas.
If $c_i$ and $c_j$, $i\neq j$, do not conflict, then requests in $c_i$ and $c_j$ can execute concurrently.
\end{itemize}

 \begin{figure*}[!ht]
  \begin{center}
  \includegraphics[width=1.6\columnwidth]{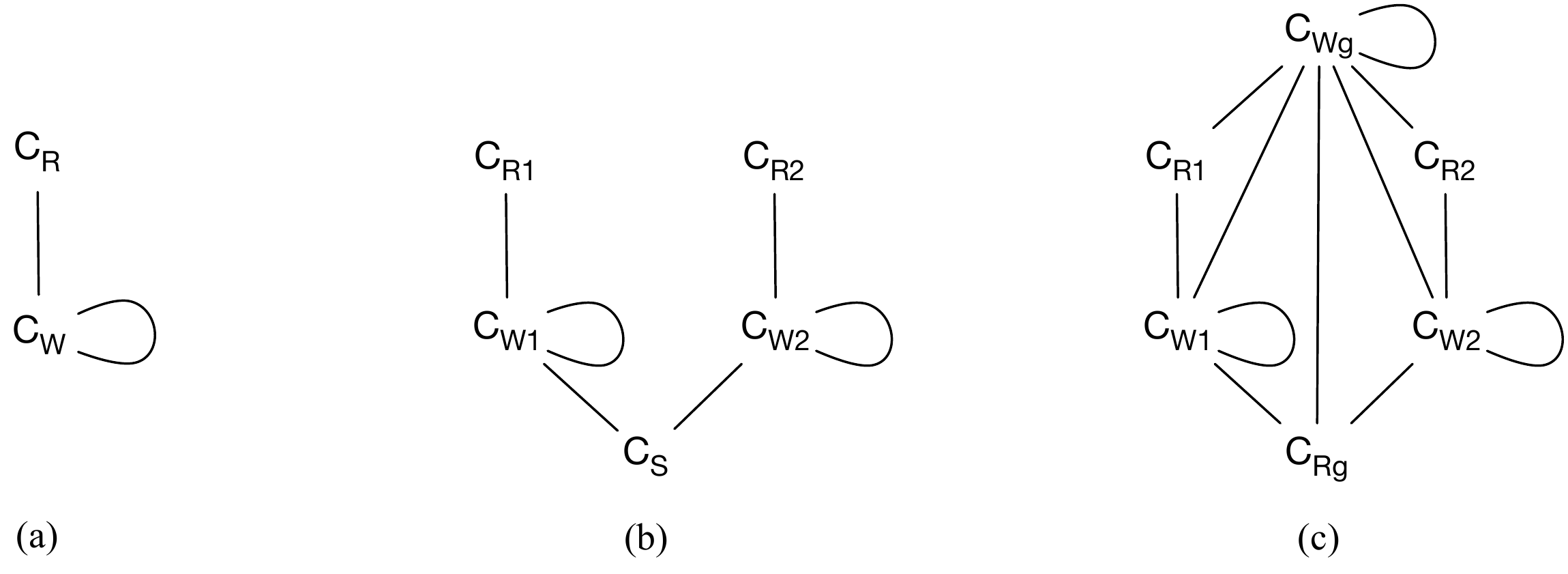}
  \end{center}
  \caption{Classes of requests for three applications: 
  (a)~readers ($C_R$) and writers ($C_W$); 
  (b)~sharded with local readers ($C_{R1}$ and $C_{R2}$), local writers ($C_{W1}$ and $C_{W2}$), and global snapshots ($C_S$); and 
  (c)~sharded with local readers and writers ($C_{R1}$, $C_{R2}$, $C_{W1}$ and $C_{W2}$), and global readers and writers $C_{Rg}$ and $C_{Wg}$. 
  Edges in the graphs represent external conflicts between classes; loops represent internal conflicts.}
  \label{fig:cclasses}
\end{figure*}

\subsection{Applications and request classes}
\label{sec:conc_samples}

We illustrate request classes with three applications.
\begin{itemize}
\item \emph{Readers and writers.} 
In this case (Fig.~\ref{fig:cclasses}~(a)), we have a class of operations $C_R$ that only read the application state and a class of operations $C_W$ that read and update the  state. 
Requests in $C_R$ do not conflict with each other (i.e., no internal conflict) but conflict with requests in $C_W$.
$C_W$ also has internal conflicts.

\item \emph{Sharded state with local readers, local writers, and global readers (snapshots).}
By sharding the service state (Fig.~\ref{fig:cclasses}~(b)), we allow concurrency within and across shards for operations that only read the state ($C_{R1}$ and $C_{R2}$) and concurrency across shards for operations that read and write the application state ($C_{W1}$ and $C_{W2}$). 
Global reads or snapshots ($C_S$) can execute concurrently with other read operations but must be synchronized with write operations.

\item \emph{Sharded state with local readers and writers and global readers and writers.}
This model (Fig.~\ref{fig:cclasses}~(c)) extends the previous one and allows both read requests ($C_{Rg}$) and write requests ($C_{Wg}$) that span multiple shards. 
\end{itemize}

\subsection{Replica execution model}
\label{sec:execmodel}


We adopt the following replica execution model, which extends the more restrictive model proposed in~\cite{Alchieri2017}: 
\begin{itemize}
\item All the replicas have $n+1$ threads: one scheduler thread and $n$ worker threads.
\item Each worker thread (in a replica) has a separate input queue and removes requests from the queue in FIFO order.
\item The scheduler delivers requests in total order and dispatches each request to one or more input queues, according to a mapping of classes to worker threads, which we define below.
\item If a request is scheduled to one worker only, then the request can be processed concurrently with requests enqueued in other worker threads.
\item If two or more worker threads have the same request $r$ enqueued, it is because $r$ depends on preceding requests assigned to these workers.
Therefore, all workers involved in $r$ must synchronize before one worker, among the ones involved, executes $r$.
\end{itemize}


The mapping of classes to threads must respect all interdependencies between classes. 
Intuitively, 
(a)~if a class has internal conflicts, then the synchronization mode of the class must be sequential; and
(b)~if two classes conflict, then they must be assigned to at least one thread in common.
The mapping of classes to threads is defined below and detailed in \S\ref{sec:mapping}.


\begin{definition}[Classes to threads, CtoT]
\label{def.requestClasses}
If $T = \{ t_0, .., t_{n-1} \}$ is the set of worker thread identifiers at a replica, where $n$ is the number of worker threads,
$\mathcal{T} = \{ Seq, Cnc \}$ is the synchronization mode of the class (i.e., sequential or concurrent), 
and $C$ is the set of class names as in Def.~\ref{def.rClasses}, then $CtoT = C \rightarrow \mathcal{T} \times \mathcal{P}(T)$.
\end{definition}

Algorithms~\ref{alg_scheduler} and \ref{alg_workers} present the execution model for the scheduler and worker threads, respectively.

\renewcommand{\algorithmiccomment}[1]{#1}
 \algsetup{
    linenosize=\scriptsize,
    linenodelimiter=)
 }


\begin{algorithm}[ht!]
\floatname{algorithm}{Algorithm}
\caption{Scheduler.}
\footnotesize{
\begin{distribalgo}[1]
\label{alg_scheduler} 

\INDENT{\textbf{variables:}}
	\STATE $queues[0,...,n-1] \leftarrow \emptyset $
	\COMMENT{\hfill // one queue per worker thread}
\ENDINDENT
 

\INDENT{\textbf{on deliver(req):}}
	\IF[\hfill // if execution is sequential]{$req.class.smode = Seq$}
		\STATE {$\forall t \in CtoT(req.classId) $}
		\COMMENT{\hfill // for each conflicting thread}
		\STATE {$ \  \  \  \  queues[t].fifoPut(req)$}
		\COMMENT{\hfill // synchronize to exec req}
	\ELSE[\hfill // else assign req to one thread in round-robin]
		\STATE $queues[next(CtoT(req.classId))].fifoPut(req)$ 
	\ENDIF
\ENDINDENT
\end{distribalgo}
 }
\end{algorithm}



\begin{algorithm}[ht!]
\floatname{algorithm}{Algorithm}
\caption{Worker threads.}
\footnotesize{
\begin{distribalgo}[1]
 \label{alg_workers} 

\INDENT{\textbf{variables:}}
	\STATE $myId \leftarrow id \in \{0,...,n-1 \} $
	\COMMENT{\hfill // thread $id$, out of $n$ threads}
	\STATE $queue[myId] \leftarrow \emptyset$
	\COMMENT{\hfill // the queue with requests for this thread}
	\STATE  $barrier[C]$
	\COMMENT{\hfill // one barrier per request class}
\ENDINDENT

 
\WHILE{true} 
	\STATE $req \leftarrow queue.fifoGet()$ 
	\COMMENT{\hfill// wait until a request is available}
	\IF[\hfill // sequential execution:]{$req.class.smode = Seq$}
		\IF[\hfill // smallest id:]{$myId = min(CtoT(req.classId))$} 
			\STATE $barrier[req.classId].await()$ 
			\COMMENT{\hfill // wait for signal}
			\STATE $exec(req)$
			\COMMENT{\hfill // execute request}
			\STATE $barrier[req.classId].await()$
			\COMMENT{\hfill // resume workers}
		\ELSE
			\STATE $barrier[req.classId].await()$
			\COMMENT{\hfill // signal worker}
			\STATE $barrier[req.classId].await()$
			\COMMENT{\hfill  // wait execution}
		\ENDIF
	\ELSE[\hfill // concurrent execution:] 
		\STATE $exec(req)$
		\COMMENT{\hfill // execute the request} 
	\ENDIF
\ENDWHILE
\end{distribalgo}
}
\end{algorithm}

Whenever a request is delivered by the atomic broadcast protocol, the scheduler (Algorithm~\ref{alg_scheduler})
assigns it to one or more worker threads. 
If a class is sequential, then all threads associated with the class receive the request to synchronize the execution (lines 4--6). 
Otherwise, requests are associated to a unique thread (line 7--8), following a round-robin policy (function $next$). 

Each worker thread (Algorithm~\ref{alg_workers}) takes one request at a time from its queue in FIFO order (line 6) and then proceeds depending on the synchronization mode of the class.
If the class is sequential, then the thread synchronizes with the other threads in the class using barriers before the request is executed (lines 8--14).
In the case of a sequential class, only one thread executes the request.
If the class is concurrent, then the thread simply executes the request (lines 15--16).

\subsection{Mapping classes to worker threads}
\label{sec:mapping}

Assuming the execution model described in the previous section, the mapping of classes to worker threads must satisfy the following rules:

\begin{itemize}

\item \emph{R.1: Every class is associated with at least one worker thread.}
This is an obvious requirement, needed to ensure that requests in every class are eventually executed.


\item \emph{R.2: If a class has internal conflicts, then it must be sequential.}
In this case, all workers associated to the class must be synchronized so that each request is executed by one worker in the same order across replicas.
Associating multiple workers to a sequential class does not help performance, but it induces important synchronization across classes.

\item \emph{R.3: If two classes conflict, at least one of them must be sequential.}
Requirement \emph{R.2} may help decide which class, between two conflicting classes, must be sequential.


\item \emph{R.4: If classes $c_1$ and $c_2$ conflict, $c_1$ is sequential, and $c_2$ is concurrent, then the set of workers associated to $c_2$ must be included in the set of workers associated to $c_1$.}
This requirement ensures that requests in $c_1$ are serialized with requests in $c_2$.

\item \emph{R.5: If classes $c_1$ and $c_2$ conflict and both $c_1$ and $c_2$ are sequential, 
it suffices that $c_1$ and $c_2$ have at least one worker in common.}
The common worker ensures that requests in the classes are serialized.

\end{itemize}

The previous rules ensure linearizable executions, the correctness proofs are in Appendix A.

\section{Optimizing scheduling}
\label{sec:optm}
\label{sec:optimization}

In this section, we formulate the problem of mapping a class of requests to a set of worker threads.
The mapping assumes the execution model described in \S\ref{sec:execmodel} and satisfies the rules presented in \S\ref{sec:mapping}.

\subsection{The optimization model}
\label{sec:themodel}

The mapping of threads to classes is shaped as an optimization problem that must produce a solution that satisfies a number of constraints and is optimized for a few conditions (see Algorithm~\ref{model}).

\paragraph{Input and output (lines \ref{lC}--\ref{ldep}).}

The input is the request classes definition, a set of available worker threads, the conflicts among classes, and the relative weight of classes.
\begin{itemize}
\item The set of classes $C = \{c_1, .., c_{nc} \}$, as in Def. \ref{def.rClasses}.
\item The set of worker threads $T = \{ t_0, .., t_{nt-1} \}$ that will be mapped to classes.
\item The class conflict relation $\# \subseteq C \times C$, derived from $\mathcal{R}$ in Def. \ref{def.RCtoR},
where $(c_1,c_2) \in \# $ iff $ (c_1 \rightarrow (c_{set},r_{set})) \in \mathcal{R}$ and $c_2 \in c_{set} $.
\item The weight of classes $W \subseteq C \times  \mathbb{R}$, used to represent the work imposed by each class.
We assign weights to classes based on the frequency of requests in the class in the workload.
(If such information is not available, we assume equal weight to all classes.)
\end{itemize}

The optimization of this model has as output the mapping of worker threads to classes and the choice of whether a class is sequential or concurrent.
\begin{itemize}
\item The mapping $uses\subseteq C \times T$ of worker threads to classes; where $uses(c,t)$ means that worker thread $t$ is mapped to class $c$.
\item The synchronization mode of each class, either sequential or concurrent. 
\end{itemize}

\paragraph{Constraints (lines \ref{constr}--\ref{lIR5}).} 

The output of the optimizer must follow the set of restrictions R.1--R.5, as defined in \S\ref{sec:mapping}.

\paragraph{Objective function (lines \ref{objectives}--\ref{lobjEnd}).} 

The restrictions imposed by the constraints may be satisfied by several associations of threads to classes. 
Among those, we aim at solutions that minimize the ``cost of execution", as defined below.
\begin{itemize}
\item \emph{O.1: Minimize threads in sequential classes and maximize threads in concurrent classes.} 
We account for this by penalizing sequential classes proportionally to the number of threads assigned to the class (\emph{O.1a}), and rewarding (i.e., negative cost) concurrent classes proportionally to the number of threads assigned to the class (\emph{O.1b}). In both cases, we also take into account the weight of the class with respect to other classes with the same synchronization mode.
\item \emph{O.2: Assign threads to concurrent classes in proportion to their relative weight.} 
More concretely, we minimize the difference between the normalized weight of each concurrent class $c$ (i.e., $w[c]/wc$) and the number of threads assigned to $c$ relative to the total number of threads (i.e., $|\{\forall t \in T:uses[c,t]\}|/ \mathit{nt}$).
\item \emph{O.3: Minimize unnecessary synchronization among sequential classes.} 
%
We penalize solutions that introduce common threads in independent sequential classes. 
Since the classes are sequential, a single thread in common would make the classes mutually exclusive.
Moreover, we give this objective higher priority than the others.
\end{itemize}

\newcommand{\spaceone}{0.8mm}
\newcommand{\spacetwo}{1.4mm}

\begin{algorithm}[ht!]
\floatname{OptModel}{Model}
\caption{Optimization model.}
\footnotesize{
\begin{distribalgo}[1]
\label{model} 

\INDENT{\textbf{input:}} \label{lC}
	\vspace{\spaceone}
	\STATE $set \ C$ 
	\COMMENT{\hfill // set of request classes}
	\vspace{\spaceone}
	\STATE $set \ T$ \label{lT}
	\COMMENT{\hfill // set of worker threads}
	\vspace{\spaceone}
	\STATE $\#\{C_i,C_j\} \in \{ 0, 1 \}$ \label{lconflicts}
	\COMMENT{\hfill // $\#[c_i,c_j] = 1$\ :\ $c_i$ and $c_j$ conflict}
	\vspace{\spaceone}
	\STATE $w\{C\} \in \mathbb{R}$ \label{lw}
	\COMMENT{\hfill // weight of each class}
	\vspace{\spaceone}
\ENDINDENT

\INDENT{\textbf{output:}}
	\vspace{\spaceone}
	\STATE $uses\{C,T\} \in \{ 0, 1 \}$ \label{luses} 
	\COMMENT{\hfill // uses[c,t] = 1 : class $c$ uses thread $t$}   
	\vspace{\spaceone}
	\STATE $Seq\{C\} \in \{ 0, 1 \}$ \label{ldep}
	\COMMENT{\hfill  // $Seq(c) = 1$ : $c$ is sequential}
	\vspace{\spaceone}
\ENDINDENT

\INDENT{\textbf{auxiliary definitions:}}
	\vspace{\spaceone}
	\STATE $nc = |C|$
	\COMMENT{\hfill  // number of classes}
	
	\vspace{\spaceone}
	\STATE $nt = |T|$
	\COMMENT{\hfill  // number of worker threads}
	
	\vspace{\spaceone}
	\STATE $Cnc\{C\} = \lnot Seq\{C\} \in \{ 0, 1 \}$ \label{ldep2}
	\COMMENT{\hfill  // $Cnc(c)\!=\!1\!:\!c$ is concurrent}

	\vspace{\spaceone}
	\STATE $wc = \sum_{\forall c \in C\ :\ Cnc[c]} w[c]$
	\COMMENT{\hfill // weights of concurrent classes}

	\vspace{\spaceone}
	\STATE $ws = \sum_{\forall c \in C\ :\ Seq[c]} w[c]$
	\COMMENT{\hfill // weights of sequential classes}

	\vspace{\spaceone}
\ENDINDENT

\INDENT{\textbf{constraints:}}\label{constr}
	\vspace{\spaceone}
	\STATE $\forall c \in C \ :\  \bigvee_{\forall t \in T} uses(c,t)$ \label{lIR0}
	\COMMENT{\hfill // R.1}
	\vspace{\spacetwo}
	
	\STATE $\forall c \in C \ :\  \#[c_1,c_1] \Rightarrow Seq[c_1]$ \label{lIR1}
	\COMMENT{\hfill // R.2}
	\vspace{\spacetwo}
	
	\STATE $\forall c_1, c_2 \in C \ :\ \#[c_1,c_2] \Rightarrow Seq[c_1] \lor Seq[c_2]$ \label{lIR3}
	\COMMENT{\hfill // R.3}
	\vspace{\spacetwo}

	\STATE $\forall c_1, c_2 \in C, t \in T \ :\ $ \\
		\hfill $\#[c_1,c_2] \land Seq[c_1] \land Cnc[c_2] \land uses[c_2,t] \Rightarrow uses[c_1,t]$\\ \label{lIR4}
 	\COMMENT{\hfill // R.4}
	\vspace{\spacetwo}

	\STATE $\forall c_1, c_2 \in C \ :\ \#[c_1,c_2] \land Seq[c_1] \land Seq[c_2] \Rightarrow $ \\
		\hfill $\exists t \in T \ :\ uses[c_1,t] \land uses[c_2,t]$ \label{lIR5}
	\COMMENT{\hfill // R.5}
	\vspace{\spaceone}
\ENDINDENT

\INDENT{\textbf{objective:}}\label{objectives}
	\vspace{\spaceone}
	\STATE minimize cost:
	\vspace{\spacetwo}

    \STATE $ + \sum_{\forall t \in T, \, \forall c \in C \,:\, Seq[c]} uses[c,t] \times w[c] \, \, / \, ws$
    \COMMENT{\hfill // O.1a}
	\vspace{\spacetwo}

    \STATE $ - \sum_{\forall t \in T, \, \forall c \in C \,:\, Cnc[c]} uses[c,t] \times w[c] \, \, / \, wc$ \label{lobjInit}
   \COMMENT{\hfill // O.1b}
	\vspace{\spacetwo}
         
	\STATE $ + \sum_{\forall c \in C \,:\, Cnc[c]} | w[c]/wc -(|\{\forall t \in T:uses[c,t]\}|/ \mathit{nt} )|$\\ \label{lobjloadb} 
	\COMMENT{\hfill // O.2}
	\vspace{\spacetwo}

	\STATE $ + \sum_{\forall c_1, c_2 \in C \,:\, Seq[c_1] \land Seq[c_2] \land \lnot \#[c_1,c_2]}$ \\
	\vspace{\spaceone}
		\hfill $| \{\forall t \in T \, :\, uses[c_1,t] \land uses[c_2,t] \} | \times nt \times nc$ \label{lobjend} \label{lobjEnd}
	\COMMENT{\hfill // O.3}
	\vspace{\spacetwo}
\ENDINDENT
      
\end{distribalgo}
}
\end{algorithm}

We described the optimization problem in the AMPL language \cite{fourer1993ampl} and solved it with the KNitro solver \cite{Byrd06knitro:an}.
The problem description in AMPL is roughly similar to Algorithm~\ref{model}. 

\section{Experimental evaluation}
\label{exp}
 


In order to compare the performance of the two scheduling methods for parallel state machine replication, we implemented the CBASE late scheduler~\cite{Kot04} and the  early scheduling technique proposed in this paper and conducted several experiments.

\subsection{Environment}

We implemented both early and late scheduling in \textsc{Bft-SMaRt}~\cite{Bes14}.
\textsc{Bft-SMaRt} can be configured to use protocols optimized to tolerate crash failures only or Byzantine failures.
In all the experiments, we configured \textsc{Bft-SMaRt} to tolerate crash failures.
\textsc{Bft-SMaRt} was developed in Java and its atomic broadcast protocol executes a sequence of consensus instances, where each instance orders a batch of requests.
To further improve the performance of the 
\textsc{Bft-SMaRt} ordering protocol, we implemented interfaces to enable clients to send a batch of requests inside the same message. 

In late scheduling, scheduler and workers synchronize access to the graph using a lock on the entire graph, as suggested in~\cite{Kot04}.
As a consequence, the performance of late scheduling is impacted by the size of the dependency graph since scheduler and workers traverse the graph to include and remove edges (dependencies), respectively. 
Providing efficient concurrent access to a graph is an open problem \cite{kallimanis16, peri16}.
In the experiments, we carefully configured the size of the dependency graph so that late scheduling produces its best performance in our setup: at most 50 entries in the graph for workloads with writes only and 150 entries otherwise.

The experimental environment was configured with 9 machines connected in a 1Gbps switched network. The software installed on the machines was CentOS Linux 7.2.1511 and a 
64-bit Java virtual machine version 1.8.0\_131. \textsc{Bft-SMaRt} was configured with 3 replicas hosted in separate machines 
(Dell PowerEdge R815 nodes equipped with four 16-core AMD Opteron 6366HE processors running at 1,8 GHz and 128 GB of RAM) 
to tolerate up to 1 replica crash, while up to 200 clients were distributed uniformly across another 4 machines (HP SE1102 nodes equipped with two quad-core Intel Xeon L5420 processors running at 2.5 GHz 
and 8 GB of RAM).

 \begin{figure*}[!ht]
 \begin{center}
  \subfigure[Read - Light]{\label{t-l}
   \includegraphics[scale=0.525]{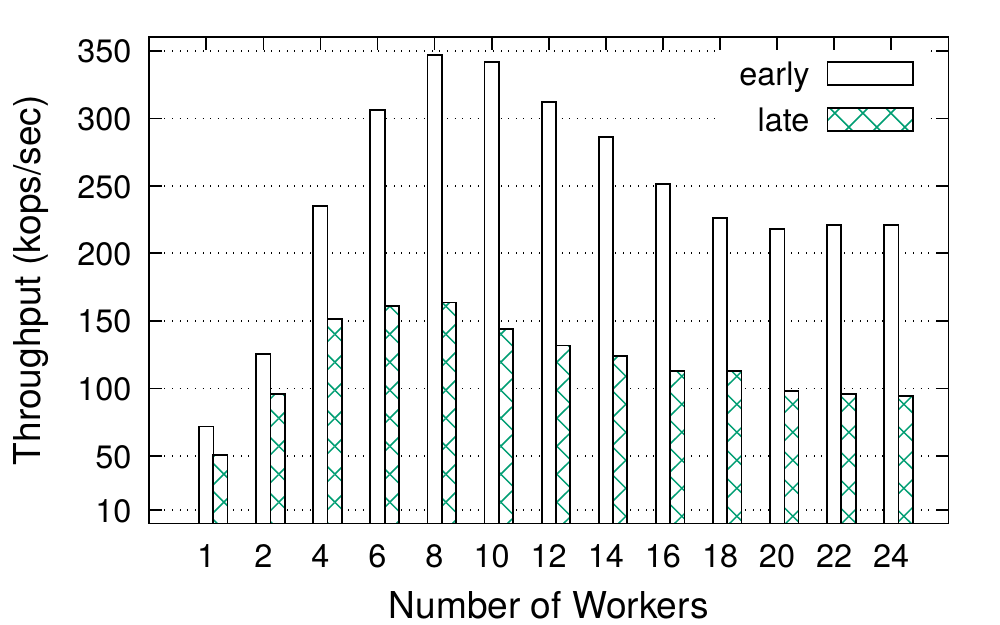}}
  \subfigure[Read - Moderate]{\label{t-m}
   \includegraphics[scale=0.525]{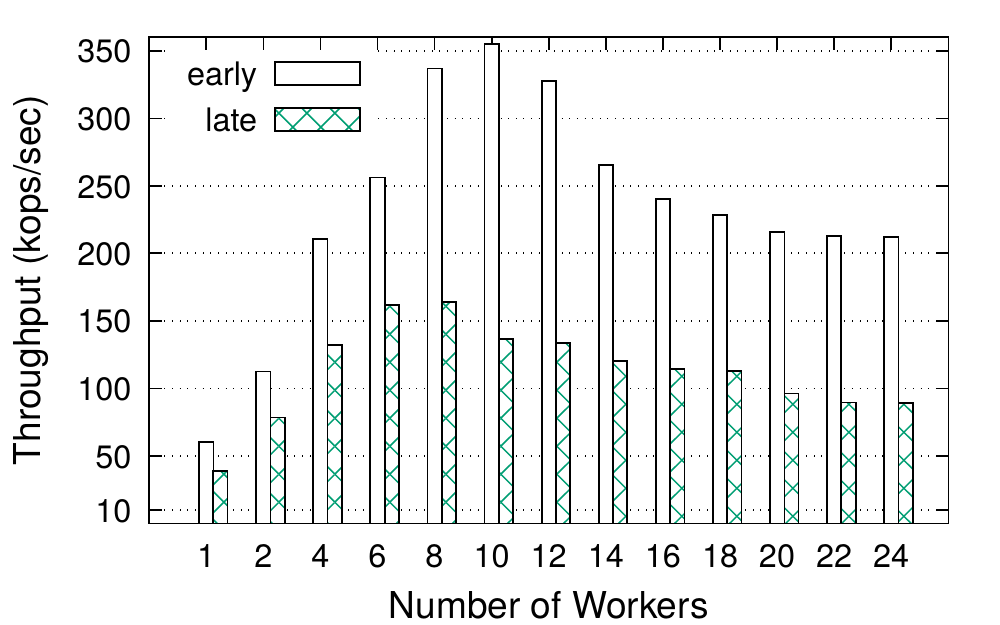}}
   \subfigure[Read - Heavy]{\label{t-h}
   \includegraphics[scale=0.525]{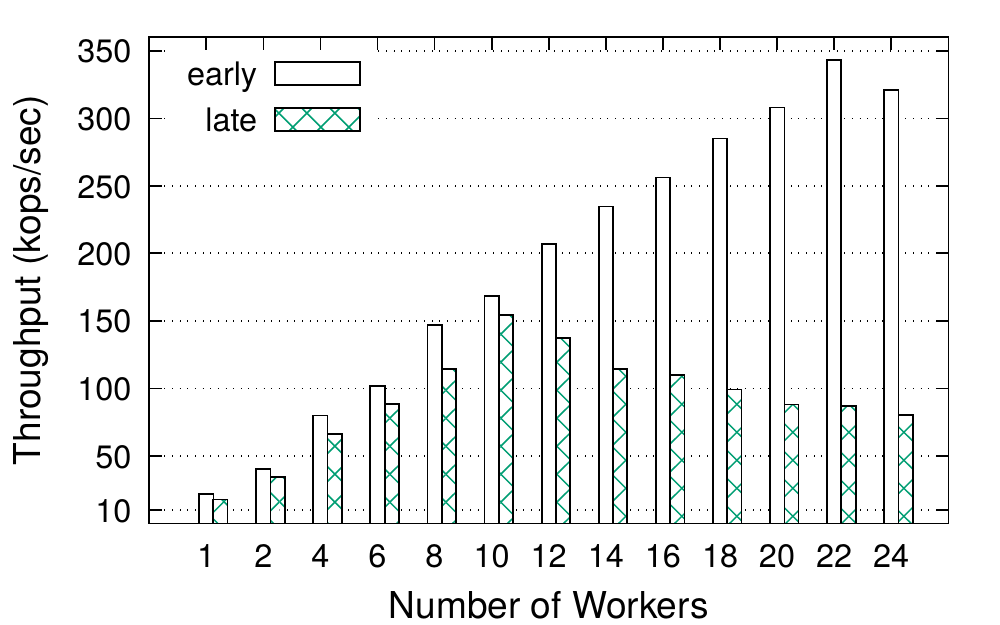}}

  \subfigure[Mixed - Light]{\label{tm-l}
     \includegraphics[scale=0.525]{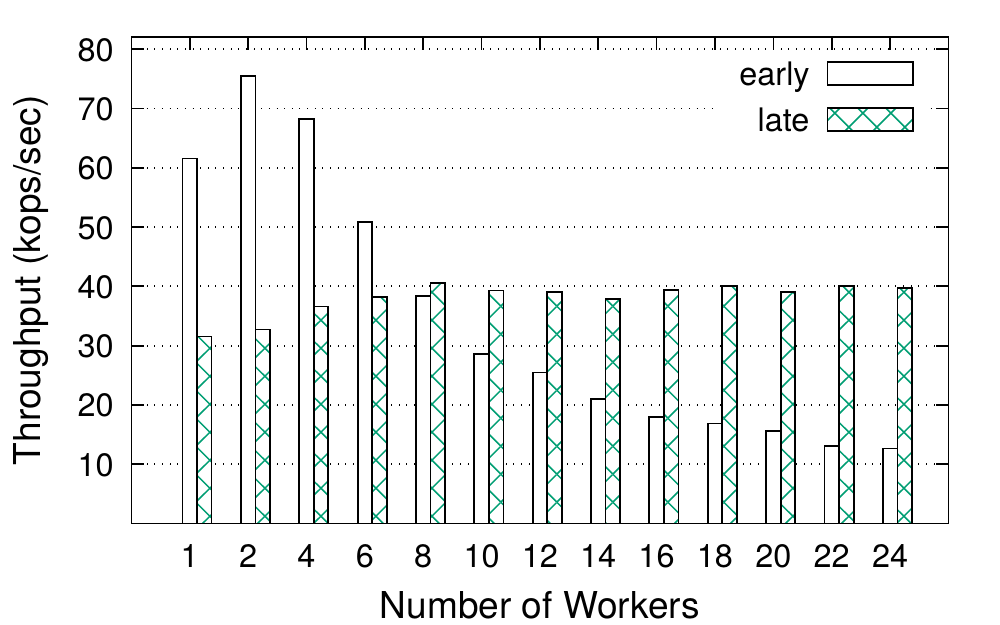}}
     \subfigure[Mixed - Moderate]{
     \label{tm-m}
     \includegraphics[scale=0.525]{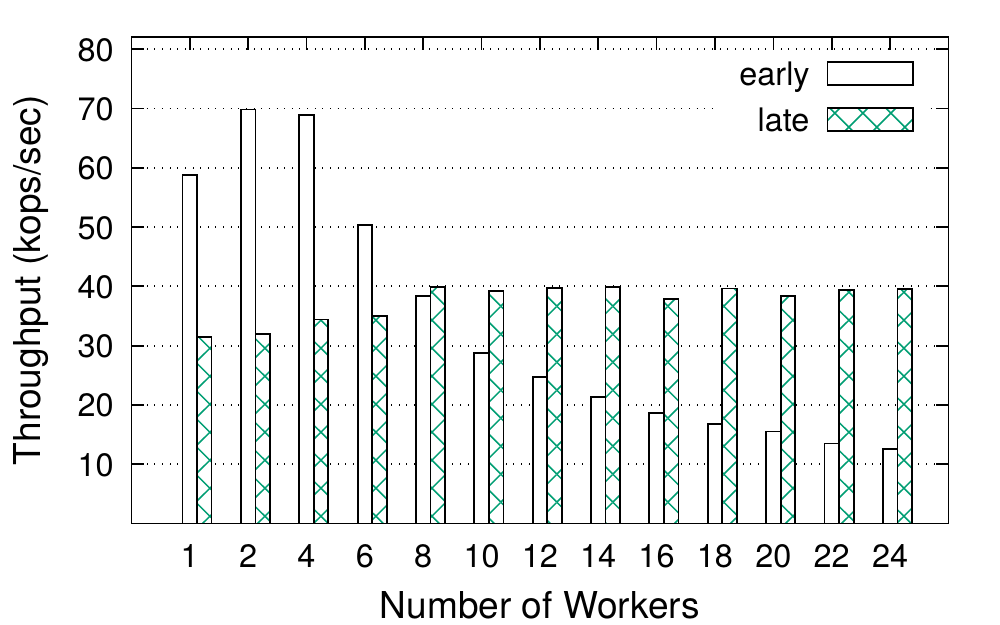}}
     \subfigure[Mixed - Heavy]{
     \label{tm-h}
     \includegraphics[scale=0.525]{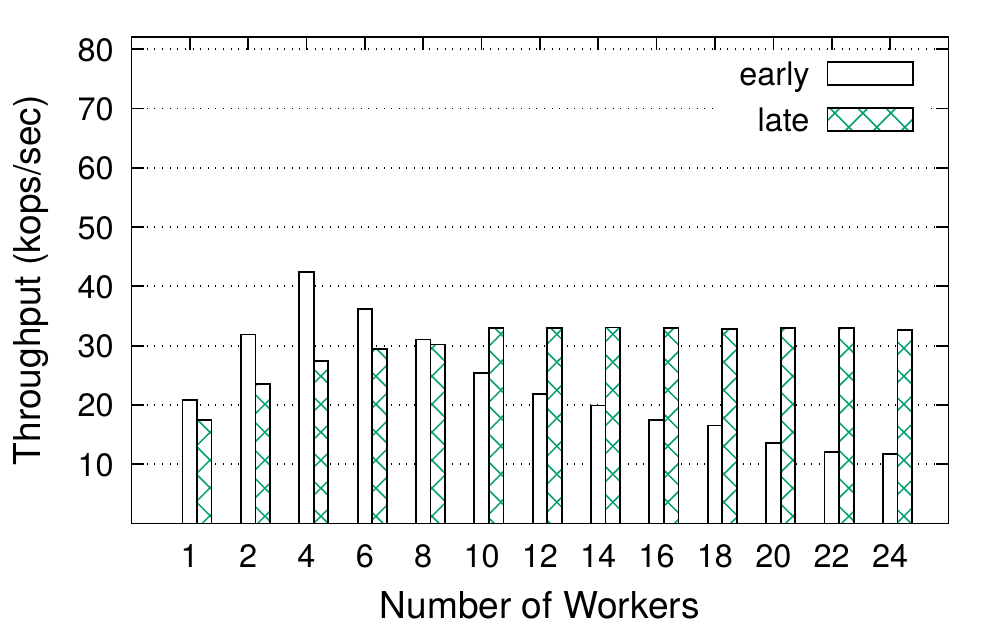}}

    \end{center}
 \caption{Throughput in readers-and-writers service (single shard) for different workloads and execution costs.}
 \label{performance_1pR}
\end{figure*}

 \begin{figure*}[!ht]
 \begin{center}
  \subfigure[Mixed - Light]{
   \includegraphics[scale=0.525]{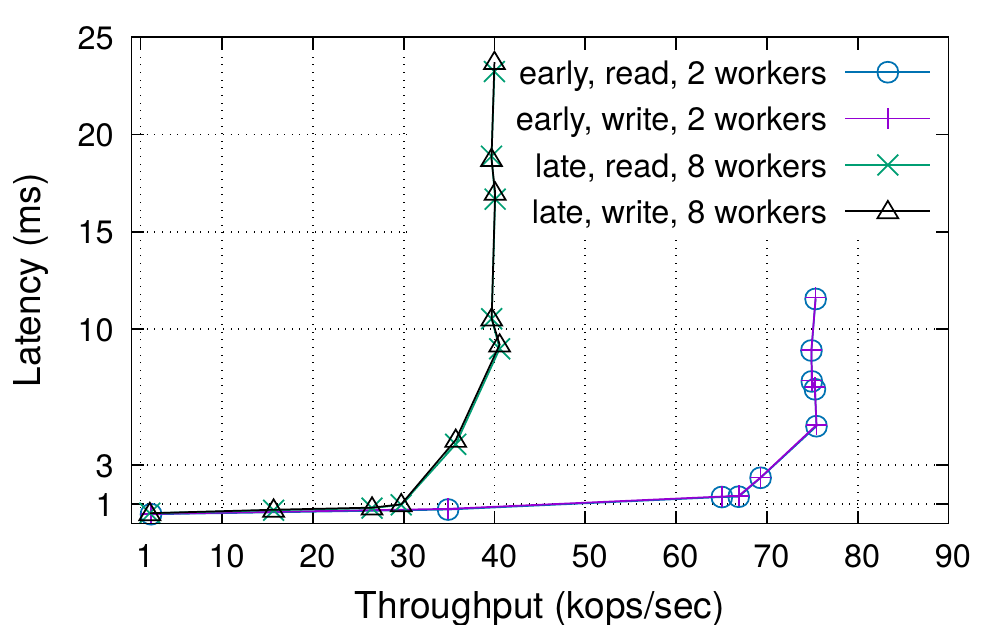}}
   \subfigure[Mixed - Moderate]{\label{l-m}
   \includegraphics[scale=0.525]{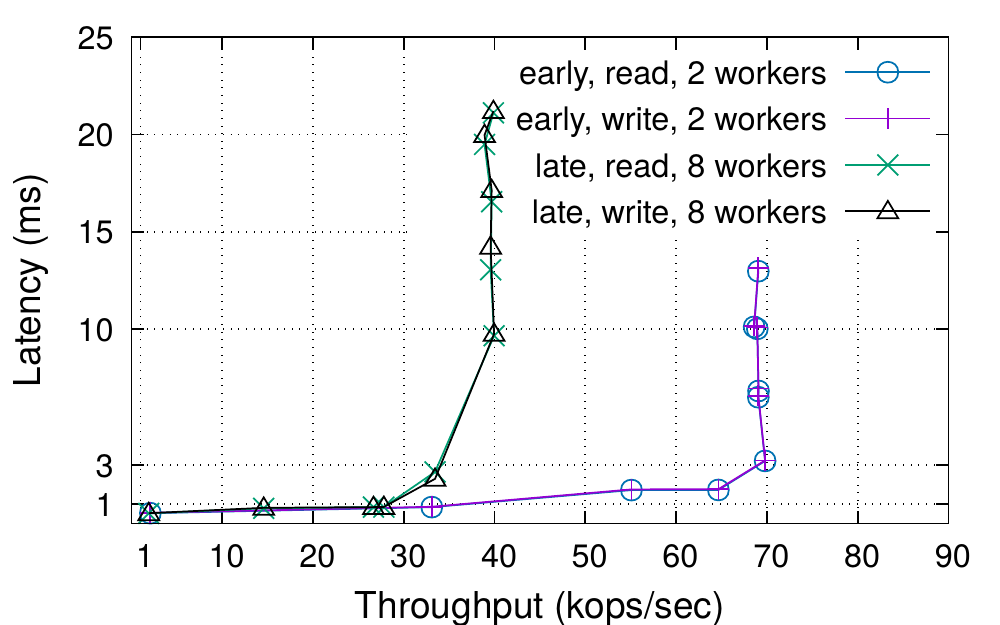}}
   \subfigure[Mixed - Heavy]{
   \includegraphics[scale=0.525]{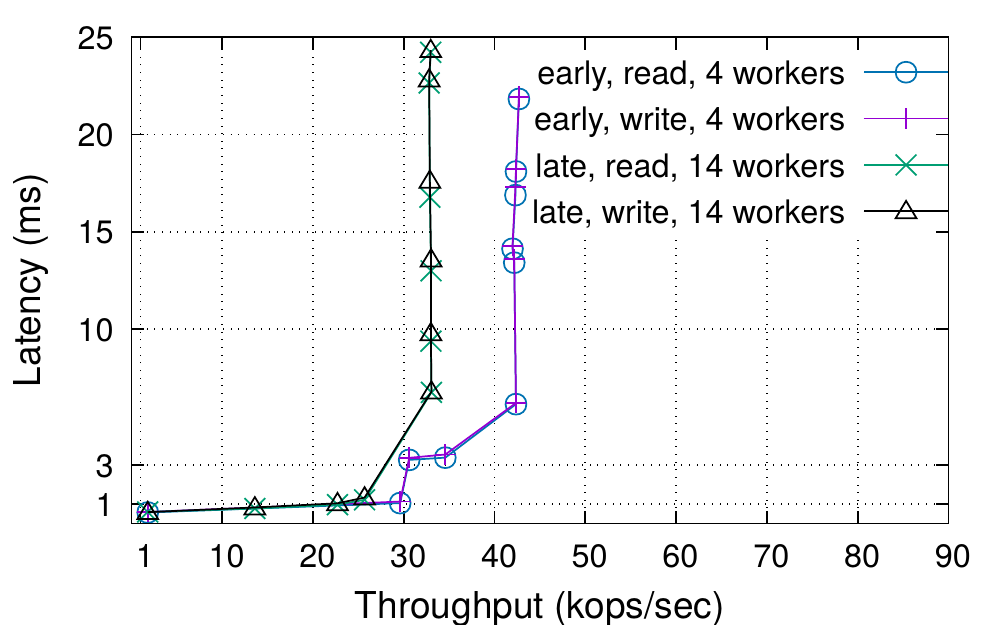}}
   
 \end{center}
 \caption{Latency in readers-and-writers service (single shard) for a mixed workload and different execution cost.}
 \label{performance_1pMix}
\end{figure*}

 %
 %
%
%
%

\subsection{Application}

The application consists of a set of linked lists. Each list stored integers and represented one shard (partition). 
For single-shard reads and writes, we support operations to verify if a list $contains$ some entry and to $add$ an item to a list: 
$contains(i)$ returns \textit{true} if $i$ is in the list, otherwise returns \textit{false};
$add(i)$ includes $i$ in the list and returns \textit{true} if $i$ is not in the list, otherwise returns \textit{false}.
Parameter $i$ in these operations was a randomly chosen position in the list.
For multi-shard reads and writes, we implemented operations 
$containsAll$ and $addAll$ to execute the respective operations involving all shards. 
Lists were initialized with $1$, $1k$ and $10k$ entries at each replica, representing operations with light, moderate, and heavy execution costs, respectively.


\subsection{Goals of the evaluation}

The main goal of our experimental evaluation is to compare the two scheduling techniques and identify in which configurations one outperforms the other.
We consider a parameter space containing
(a)~two prototypical applications with one or more shards, single-shard and multi-shard read and write operations (i.e., cases 1 and 3 presented in Fig.~\ref{fig:cclasses});
(b)~three execution costs, comprising light, moderate and heavy operations; and
(c)~uniform and skewed workloads (i.e., the three cases depicted in Fig.~\ref{fig:classthreads}).
We evaluated the throughput of the system at the servers and the latency perceived by the clients, a warm-up phase preceded each experiment.


\subsection{Single-sharded systems} 
\label{ex:1p}

  


The first set of experiments considers reads and writes in a single shard.
Fig.~\ref{performance_1pR} shows the throughput presented by each scheduling approach for different execution costs and number of workers threads, considering two workloads: 
a workload with read operations only and a mixed workload composed of $15$\% of writes and $85$\% of reads uniformly distributed among clients.

Considering the read operations only workload (Figs.~\ref{t-l}, \ref{t-m} and \ref{t-h}), in general the early scheduler presented more than twice the performance of the late scheduler. 
The late scheduler uses one data structure (dependency graph) that is accessed by the scheduler to insert requests and by the workers to remove requests, leading to contention. 
The shared dependency graph rapidly becomes a bottleneck and performance does not increase with additional workers. Another important aspect is that the late scheduler presented similar performance 
for all execution costs, indicating that its performance was bounded by the synchronization and contention in the dependency graph and not by the time the workers spent to execute a request.

Following a different approach, the early scheduler employs one data structure (queue) per worker.
Consequently, workers do not content with one another, although the scheduler and each worker share a data structure.
For each execution cost the system reached peak throughput with a different number of threads: for light, moderate and heavy requests it was necessary $8$, $10$ and $22$ workers, respectively. 
Early scheduling largely outperforms late scheduling since in the former there is no synchronization among workers. 

A final remark about these experiments is that after the peak throughput is reached, increasing the number of workers hampers performance.
Considering the early scheduler, this happens for the following reason.
After executing a request, a worker thread checks its queue for requests and if no request is found it blocks until one is available.
As the number of threads increases, there are fewer requests enqueued per worker, which finds its queue empty more often.
Considering a run of the experiment with light execution cost (Fig.~\ref{t-l}), $201.543$ system calls to sleep (block) are executed when the system is configured 
with $8$ workers (an average of $25.192$ system calls per worker) and $31.016.088$ such calls are executed in the configuration with $24$ workers (an average of $1.292.337$ system calls per worker). 
Although presenting a lower impact, this phenomenon also occur in the late scheduler.

Figs.~\ref{tm-l}, \ref{tm-m} and \ref{tm-h} show throughput results for the mixed workload. 
For each execution cost, Fig.~\ref{performance_1pMix} also shows latency versus throughput results for the thread configuration that presented the best performance in each scheduling approach. 
The early scheduler technique again presented the best peak throughput, which was achieved with $2$, $2$ and $4$ workers for light, moderante and heavy operations costs, respectively. 
This configuration represents an equilibrium between the synchronization needed to execute writes and the number of workers available to execute 
parallel reads. Reads and writes have similar latency because they have similar execution costs and the synchronization of writes impacts the performance of reads ordered after a write.

In workloads with predominance of write operations, sequential execution presents best performance~\cite{Alchieri2017,Par14}. 
For example, considering a workload with only write operations of moderante costs, the early scheduling presented a throughput of 53 Kops/sec with 1 worker that dropped down to 24 kops/sec with 2 workers. 
Notice this is the worst case for this approach, where all workers need to synchronize for each request execution. 
The late scheduling presented a throughput of 31 Kops/sec with 1 worker and 23 kops/sec with 2 workers. 
For an intermediary workload composed of $50$\% of reads and $50$\% of writes, the early scheduling presented a throughput of 55 kops/sec with 1 worker and 41 kops/sec with 2 workers. The late scheduling 
presented a throughput of 39 Kops/sec with 1 worker and 36 kops/sec with 2 workers.


\subsection{Optimizing multi-sharded systems}
\label{sec:example}

\begin{figure}[!ht]
  \begin{center}
  \includegraphics[width=0.99\columnwidth]{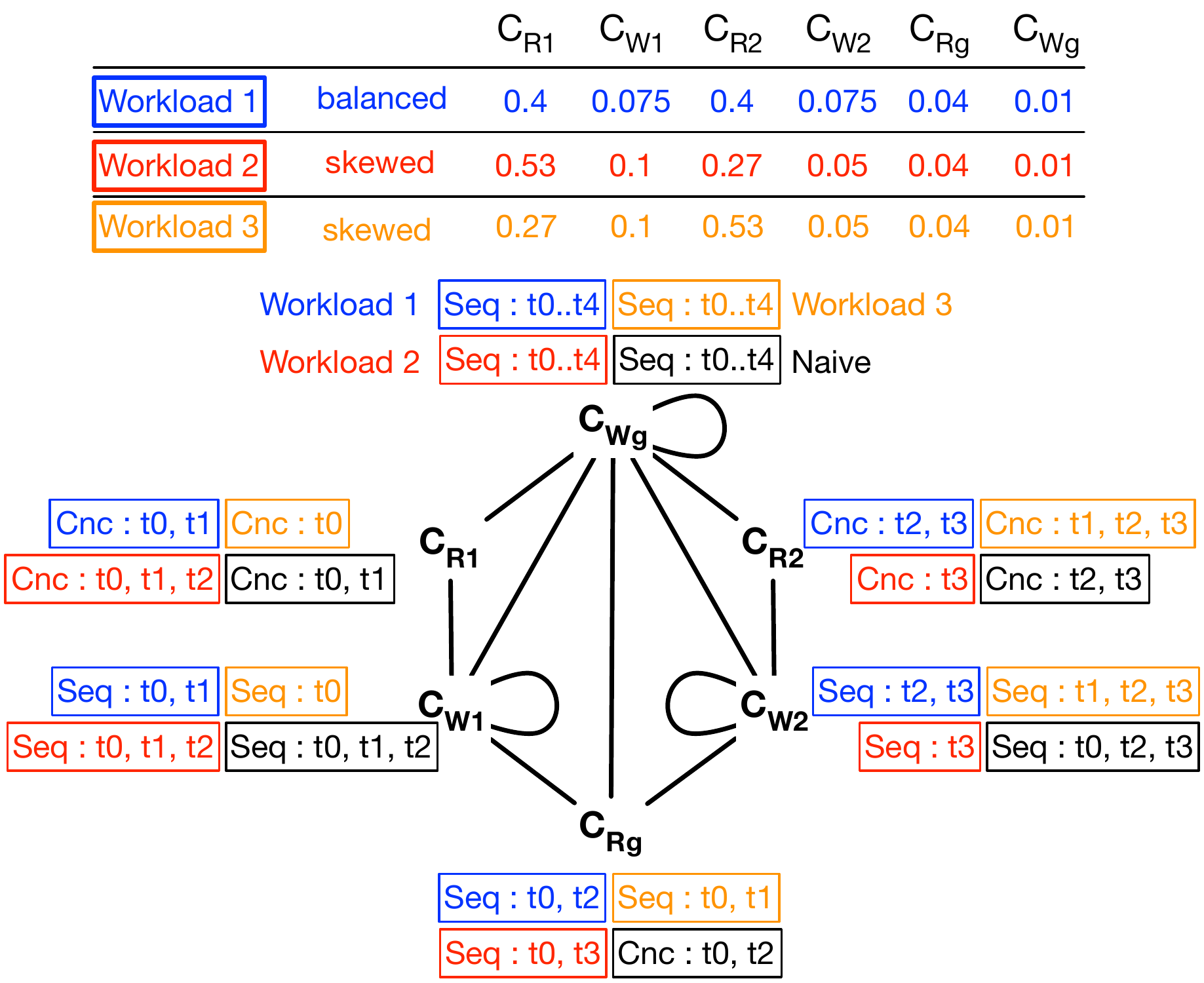}
  \end{center}
  \caption{Thread assignment in sharded application.}
  \label{fig:classthreads}
\end{figure}

We now evaluate our optimization model.
We consider the generalized sharded service depicted in Fig. \ref{fig:cclasses}(c) with 2 shards, 4 worker threads, and three workloads, as described below.
For each workload, Fig.~\ref{fig:classthreads} shows the resulting proportion of local reads, local writes, global reads, and global writes.

\begin{itemize}
\item \emph{Workload 1: Balanced load.}
This workload is composed by 85\% of reads and 15\% of writes, from which 95\% are single-shard (local), equally distributed between the two shards, and 5\% are multi-shard (global). 


\item \emph{Workload 2: Skewed load, shard 1 receives more requests than shard 2.}
Similar to \textit{Workload 1}, but local requests are unbalanced with 67\% to shard 1 and 33\% to shard 2. 

\item \emph{Workload 3: Skewed load, shard 1 receives more writes and fewer reads than shard 2.}
Similar to \textit{Workload 1}, but local requests are unbalanced with 67\% of writes and 33\% of reads to shard 1, and 33\% of writes and 67\% of reads to shard 2. 
\end{itemize}

Fig.~\ref{fig:classthreads} shows the results produced by the optimizer for each workload and by a naive assignment of threads.
In the naive assignment, we first configure read-only classes to be concurrent and assign workers more or less proportionally to the  percentage of commands in the workload.
We then assign workers to write classes to ensure the correctness of the model.

For all workloads, our optimization model establishes that reads and writes in shards 1 and 2 use disjoint sets of threads, and threads are distributed in proportion to the class weights.   
Since single-shard reads are concurrent but conflict with writes in the shard, the single-shard write class has all threads associated to the respective read class. 
Multi-shard writes must be synchronized with all other operations, therefore all threads are assigned to multi-shard writes. 
According to the conflict definition, multi-shard reads are concurrent.
Nevertheless, the optimizer sets the class $C_{Rg}$ as sequential.
The reason is that since $C_{Rg}$ conflicts with $C_{W1}$ and $C_{W2}$, if $C_{Rg}$ is configured as concurrent all threads assigned to $C_{Rg}$ would have to be included in $C_{W1}$ and $C_{W2}$.
Doing so would synchronize $C_{W1}$ and $C_{W2}$ since their threads would not be disjoint.
A more efficient solution is to define $C_{Rg}$ as sequential and assign to it one thread from $C_{W1}$ and one thread from $C_{W2}$.
As a result, multi-shard reads synchronize with local-shard writes, but local writes to different shards can execute concurrently.

\subsection{Multi-sharded systems}
\label{ex:MP}

  \begin{figure*}[!ht]
 \begin{center}
  
   \subfigure[Throughput]{\label{mp-mix}
   \includegraphics[scale=0.525]{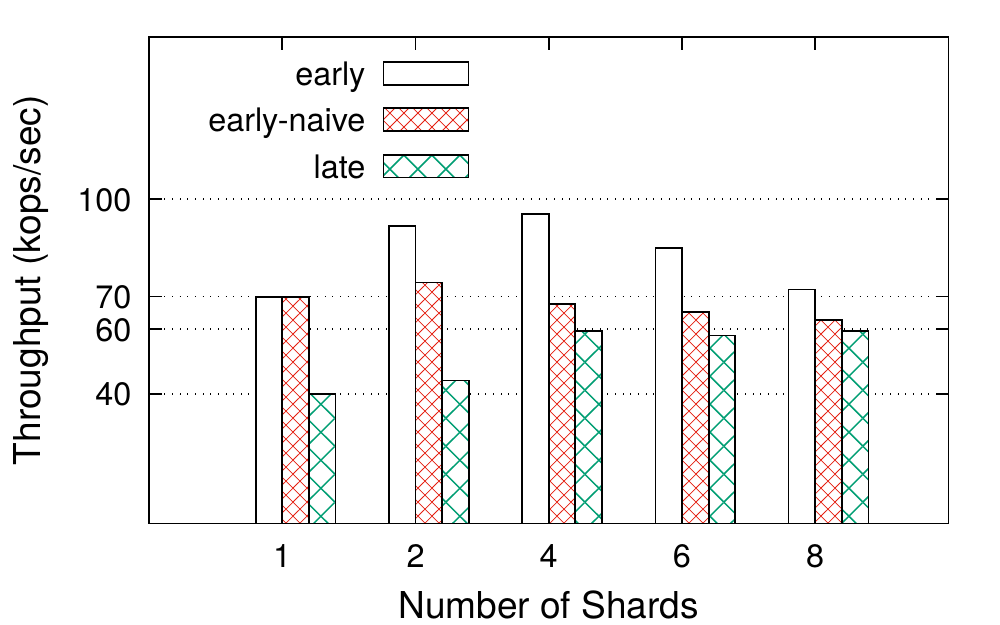}} 
  \subfigure[4 shards - single-shard - 8 workers]{\label{l-mpMixL}
   \includegraphics[scale=0.525]{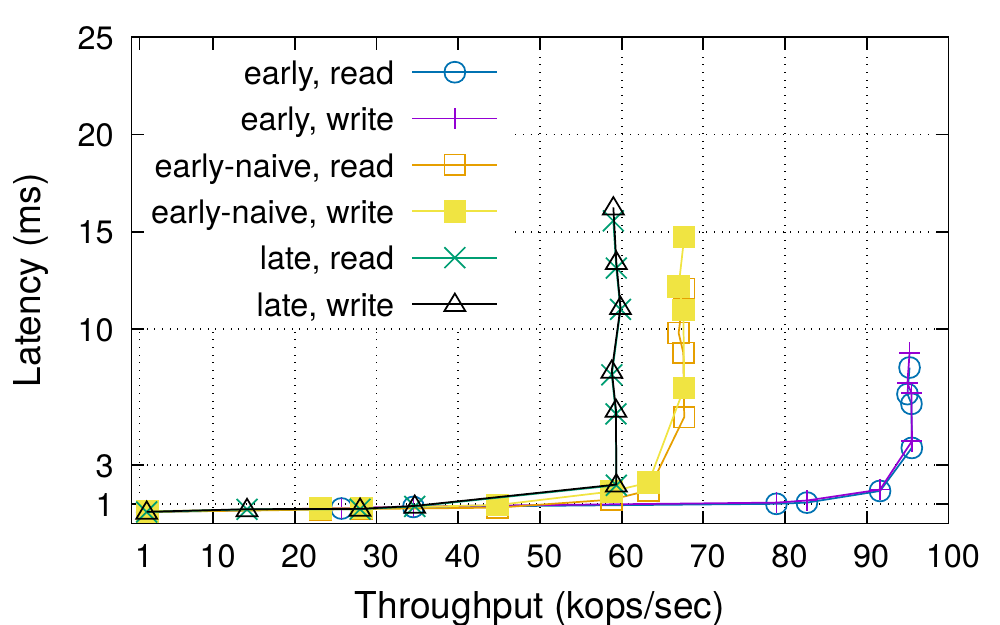}}
   \subfigure[4 shards - multi-shard - 8 workers]{\label{l-mpMixG}
   \includegraphics[scale=0.525]{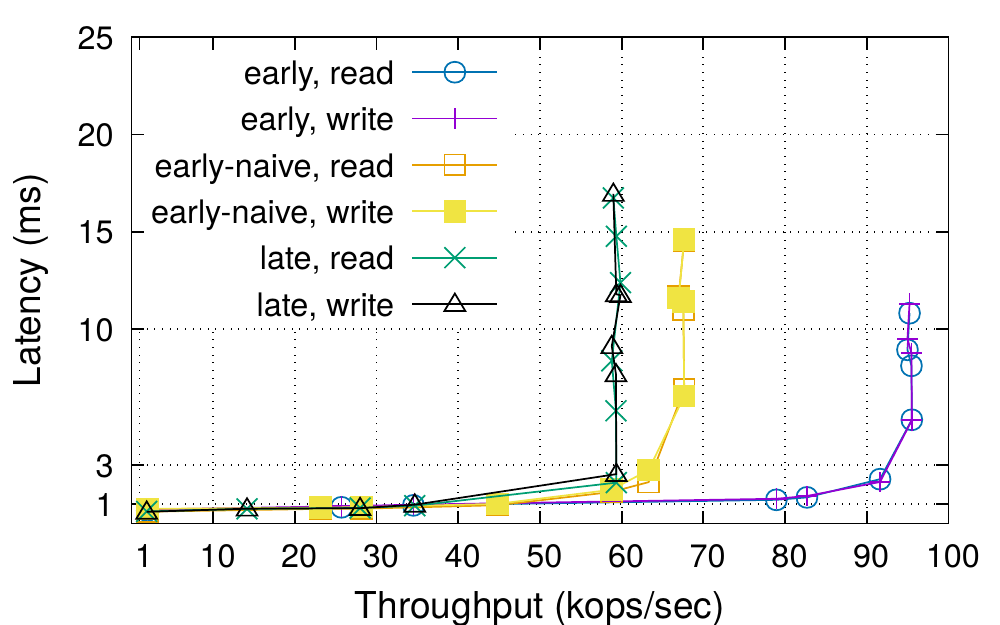}}

 \end{center}
 \caption{Multi-sharded system for a mixed workload composed of operations of moderate execution cost.}
 \label{performance_MP}
\end{figure*}

  

  


\begin{figure*}[!ht]
 \begin{center}
  \subfigure[Throughput - 4 workers]{\label{mpu-r}
   \includegraphics[scale=0.525]{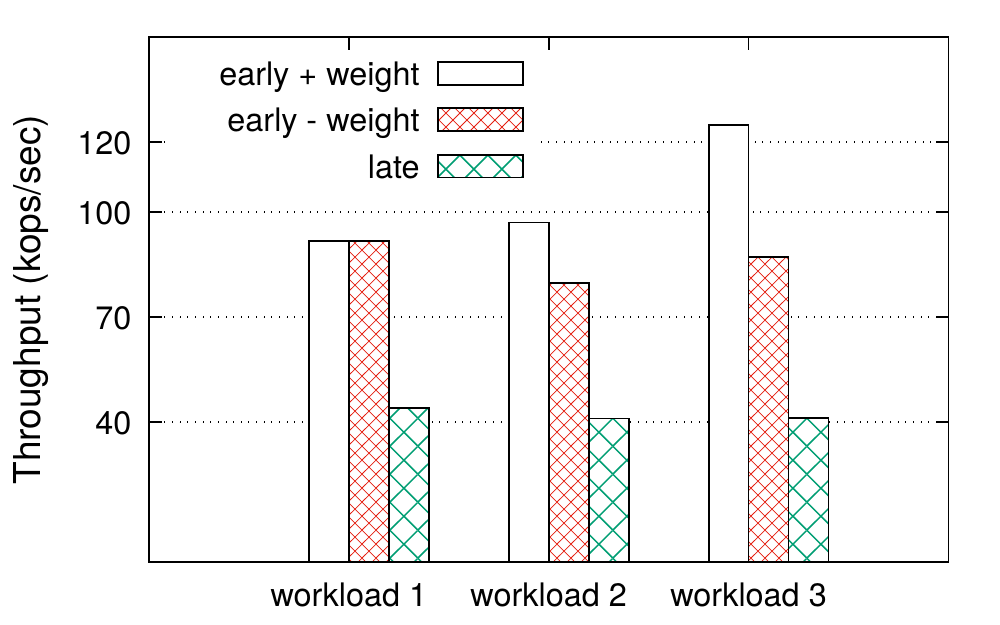}}
  \subfigure[Workload 2 - single-shard - 4 workers]{\label{l-w2s}
   \includegraphics[scale=0.525]{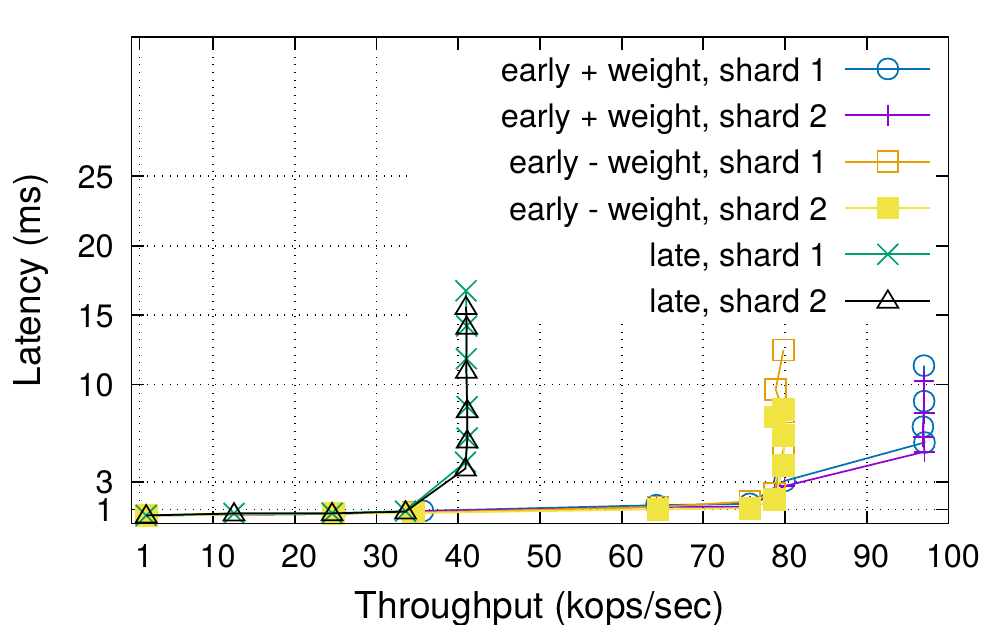}}
   \subfigure[Workload 2 - multi-shard - 4 workers]{\label{l-w2g}
   \includegraphics[scale=0.525]{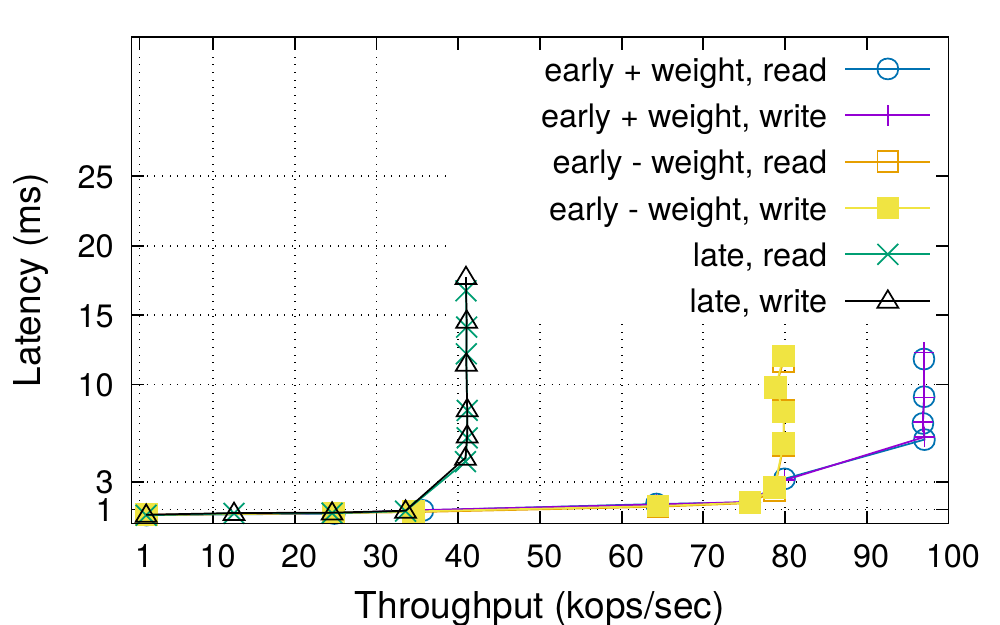}}
 \end{center}
 \caption{System with $2$ shards and skewed mixed workloads composed of operations of moderate execution cost.}
 \label{performance_UN}
\end{figure*}

The first set of experiments showed that the early scheduler outperforms the late scheduler in a single-sharded system. 
A natural question is whether this is also true for a multi-sharded system since the early scheduler statically allocates workers to shards while the late scheduler allocates workers to shards dynamically. 
Fig.~\ref{performance_MP} answers this question positively and shows the advantage of early scheduling over late scheduling for a mixed workload in systems with $1$, $2$, $4$, $6$ and $8$ shards 
and moderate operation costs. 
The workload is similar to the balanced \textit{Workload 1} introduced at \S\ref{sec:example}, extrapolated for systems with more shards, i.e., 
the single-shard operations are equally distributed among all shards. 
Moreover, operations are uniformly distributed among the clients. 

Fig.~\ref{performance_MP} also presents results for the naive threads assignment in the early scheduler in order to 
assess the impact of the proposed optimization model. Based on the results reported in Fig.~\ref{tm-m}, we configured the early scheduler with a number of threads twice the number of shards (also for the naive configuration) and the late scheduler with 8 workers in all cases.


In general, the early scheduler outperforms the late scheduler in all configurations. Moreover, the early scheduler configured with the output of the optimization model outperforms the naive configuration 
since the last does not allow parallel execution of writes assigned to different shards. The best performance was achieved with $4$ shards, suggesting that this is the best sharding for this application and 
workload, i.e., it is better to use 4 shards even if the application allows the state division in more shards. This happens because multi-shards operations need to synchronize all threads and increasing the number of shards 
also increases the number of threads involved in the synchronization. The same thread can execute operations assigned to two different shards, 
but the effect is the same as if the two shards are merged in just one. The latency evolves similarly for single- and multi-shards operations, presenting some difference only near system saturation.

\subsection{Skewed workloads}
\label{ex:UW}

We now consider workloads in which some shards receive more requests than others.
We account for skews in the workload by assigning weights to classes of requests (see \S\ref{sec:themodel}).
The weight of a class represents the frequency of class requests.
The result is that the optimizer will attempt to assign threads to classes in proportion to their weights.

Fig.~\ref{performance_UN} presents performance results for the three mixed workloads introduced at \S\ref{sec:example}, considering a system with two shards and moderate operation costs. 
Fig.~\ref{fig:classthreads} shows these workloads and how the optimizer mapped worker threads to classes. 
%
%
Fig.~\ref{performance_UN} also presents results for the early scheduler without accounting for class weights in order to better understand the effects of the technique. 
In general, the early scheduler outperforms the late scheduler even without accounting for weights. 
However, by taking into account the weights of classes performance improves since workers are mapped to classes according to the workload.

Considering \textit{Workload 2}, Fig.~\ref{l-w2s} presents the latency per shard since operations in the same shard behave similarly (see Fig.~\ref{performance_1pMix}) and 
Fig.~\ref{l-w2g} presents the latency for multi-shards operations. For early scheduling, in Fig.~\ref{l-w2s} it is possible to notice a different latency for operations addressed to different shards, mainly near system saturation. 
In the late scheduler the latency evolves similarly for both shards since the dependency graph becomes full of operations addressed to shard 1 also impacting the performance of operations in shard 2. 
Increasing the size of the dependency graph does not help since (1) larger graphs have more overhead due to the addition and removal of requests in the graph (e.g., for a graph with 500 entries, 
the write throughput is approximately 7 kops/sec) and (2) given enough time and requests, the graph will become full of such requests anyway. 
This is a serious drawback of late scheduling: given enough requests, the overall performance is bounded by operations with worst performance. 

Building on the observation above, we conducted an additional experiment in a system with 4 workers, where 50 clients read from shard 1 and another 50 clients write in shard 2. 
While early scheduler used one thread for shard 2 and the others for shard 1, executing 179 kops/sec, the late scheduler achieved only 22 kops/sec.

\subsection{Lessons learned}
 
 This section presents a few observations about the experimental evaluation.
 
\begin{enumerate}
\item \emph{Early scheduling outperforms late scheduling.}
Despite the fact that early scheduling is more restrictive than late scheduling, it delivers superior performance in almost all cases we considered.
This happens for two reasons:
(1) the early scheduler can make simple and fast decisions, and 
(2) by using one data structure per worker thread, worker threads do not content of a single data structure, as with late scheduling. 
As we could notice, in high throughput systems there is room to discuss 
the tradeoff of fast versus accurate runtime decisions. Although CBASE does not hinder
the concurrency level (since the dependencies followed are the needed and sufficient ones 
given by service semantics), better throughput results can be achieved with a 
solution trading concurrency for fast runtime decision. The main reason for this is related
to contention due to the graph structure but it is also relevant to observe that 
since the number of threads (cores) for an application is typically bounded in concrete
implementations, adopting strategies that limit the achievable concurrency level
may still not hinder the concurrency level in concrete settings.

\item \emph{Choosing the right data structures is fundamental to performance.}
Since all requests must go through the same data structure, the performance of late scheduler is bounded by requests with worst performance.
The early scheduler uses a queue per worker and does not force workers to synchronize in the presence of independent requests.
Moreover, requests of one class do not impact performance of requests in other classes.

\item \emph{The number of workers matters for performance, but not equally for the two scheduling techniques.}
For both schedulers, the ideal number of worker threads depends on the application and workload.
However, the performance impact caused by the number of workers is lower in the late than in the early scheduler.
This observation suggests that both scheduling techniques can benefit from a reconfiguration protocol~\cite{Alchieri2017} to adapt the number of workers on-the-fly.


\item \emph{Class to thread assignment needs information about all classes.}
A local view of related classes was not sufficient to reach good (in terms of the optimization objective) assignments.
The naive thread assignment shows this.  While it may seem acceptable to configure concurrent classes
distributing threads proportionally to their demands and then assure that sequential classes synchronize with
the other classes as needed, it has been observed that analysing the whole conflict relation 
could be worthy to consider a class as sequential (even it could be concurrent) to avoid further 
inducing synchronization with other classes.  
Whether this assignment problem can be partitioned (according to the class conflict topology) 
is still to be analysed.

\end{enumerate}

\section{Related Work}
\label{rw}

This paper is at the intersection of two areas of research: scheduling and state machine replication.

\subsection{Scheduling}

The area of scheduling has been active for several decades, with a rich body of contributions.
According to \cite{Leung:2004}, one classification of scheduling algorithms regards a priori
knowledge of the complete instance (i.e., set of jobs to schedule): while with \emph{offline} algorithms 
scheduling decisions are made based on 
the knowledge of the entire input instance that is to be scheduled, with \emph{online} 
algorithms scheduling decisions are taken while jobs arrive. 
Online scheduling may be further considered stochastic or deterministic.
The first arises when problem variables such as the
release times of jobs and their processing times are assumed from known distributions.
Processing times may also be correlated or independent.      
Deterministic scheduling does not consider such information. 
Several algorithms of interest are called nonclairvoyant in the sense that 
further information about the behavior of jobs (such as their processing times) 
are not known even upon their arrival (but only after completion).
Another known classification is whether the scheduling is real-time or not.
The primary concern of real-time scheduling is to meet hard deadline application 
constraints, that have to be clearly stated, while non-real-time have no such constraints.

In this context, our scheduling problem can be classified as online, deterministic (non-stochastic), and non-realtime.
Moreover, our scheduling problem imposes dependencies among jobs that have to be respected.
This aspect has also been regarded in the scheduling literature, where we can find early discussions
on how to process jobs whose interdependencies assume the topology of a directed acyclic graph, which is also the case here.
However, as far as we could survey, there is no discussion in the scheduling literature on the costs or techniques to 
detect and represent dependencies among jobs (i.e., the job dependency graph is typically given). 
Also, there is no discussion on synchronization costs to enforce job dependencies upon execution. 
While this is valid to several scheduling problems, these aspects matter when scheduling computational jobs. 
This is important in our work, as well as in related work in the area of parallel approaches to state machine replication.   
In modern computational systems, due to high throughput and concurrency,
the overhead to manage dependencies gains in importance and may become a bottleneck in the system.

\subsection{State machine replication}

State machine replication (SMR) is a well studied and central approach to the provision of fault-tolerant services.
As current and emerging applications require both high availability and performance, the 
seminal, sequential design of SMR is challenged in a number of ways.   

The sequential execution approach is followed by most SMR systems to ensure deterministic execution and thus replica consistency,
but creates an important bottleneck for applications with increasing throughput demands.   The proliferation of 
multi-core systems also calls for new designs to allow concurrent execution.  
As it has been early observed~\cite{Sch90}, independent requests can be executed concurrently in SMR.
Moreover, previous works have shown that many workloads are dominated by independent requests, which justifies strategies for concurrent request execution (e.g., \cite{Ped14, Ped16, Davi16, Kot04, Par14, Par14Opt}).  
As we briefly survey existing proposals, we observe that 
they differ in the strategy and architecture to detect and handle conflicts on the request processing path.

In \cite{Kot04}, the authors present CBASE, a parallel SMR where replicas are augmented with a deterministic scheduler.
Based on application semantics, the scheduler serializes the execution of conflicting requests according to the delivery order and dispatches non-conflicting requests to be processed in parallel by a pool of worker threads. 
Conflict detection is done at the replica side, based on pairwise comparison of requests 
in different levels of detail. Requests are then organized in a dependency graph and processed as soon as possible (i.e., whenever dependencies have been executed).  
CBASE is an example of late scheduling and studied in detail in the previous sections.

While CBASE centralizes the conflict detection and handling at the scheduler, after ordering and before executing requests,
a distinct approach is followed by Rex~\cite{guo2014rex} and CRANE \cite{cui2015paxos}, which add complexity to the execution phase.
Instead of processing conflict information to allow concurrent execution of independent requests, and synchronize conflicting requests,
both Rex and CRANE solve the non-determinism due to concurrency during request execution.
Rex logs dependencies among requests during execution, based on shared variables locked
by each request. This creates a trace of dependencies which is proposed for agreement with other follower 
replicas.
After agreement replicas replay the execution restricted to the trace of the first executing server.
CRANE \cite{cui2015paxos} implements a parallel SMR 
and solves non-determinism first by relying on Paxos and then enforcing deterministic logical times across replicas 
that combined with deterministic multi-threading~\cite{olszewski2009kendo}  allow to deterministically order
requests consistently among replicas.  

While CBASE handles conflicts before execution, Rex and CRANE during execution, Eve~\cite{kapritsos2012all} handles conflicts after execution. 
According to this approach, replicas optimistically execute requests as they arrive and check after execution if consistency is violated.
In such case, synchronization measures are taken to circumvent this problem.
In Eve replicas speculatively execute batched commands in parallel. 
After the execution of a batch, the verification stage checks the validity of replica's state.
If too many replicas diverge, replicas roll back to the last verified state and re-execute the commands sequentially.

In Storyboard \cite{kapitza2010storyboard}, a forecasting mechanism predicts the same ordered sequence of locks across replicas.
While forecasts are correct, requests can be executed in parallel.
If the forecast made by the predictor does not match the execution path of a request, 
then the replica has to establish a deterministic execution order in cooperation with the other replicas. 
  
P-SMR~\cite{Par14} avoids a central parallelizer or scheduler, to keep request execution free from additional 
overhead and without need for a posteriori checking of the execution.  This is achieved by mapping requests to 
different multicast groups.  Non-conflicting requests are sent through different multiple multicast groups that partially order
requests across replicas.  Requests are delivered by multiple worker threads according to the multicast group.
This approach imposes a choice of destination group at the client side, based on request information which is application specific.
Non-conflicting requests can be sent to distinct groups, while conflicting ones are sent to the same group(s). At replica side each worker thread is associated to a multicast group and processes requests
as they arrive.


More than proposing a strategy for parallel SMR, our approach formalizes an abstraction to represent 
classes of conflicts among requests.  This abstraction allows to express the conflict semantics of an application
with simple and common elements. Nodes are classes, edges are conflicts and requests are grouped in disjoint classes.  
This representation of conflicts and their interdependencies could be used in other contexts.
For example, in CBASE \cite{Kot04} to improve the request conflict checking or in P-SMR~\cite{Par14} to help the decision at client 
side of which multicast group to use.
\section{Conclusions}
\label{sec:conc}

This paper reports on our efforts to increase the performance of parallel state machine replication through a scheduling approach that simplifies the work done by the scheduler. 
The main idea is that by assigning requests to a worker thread (or a set of them) prior to their ordering, the scheduler is less likely to become a performance bottleneck. 
This work also proposed an optimization model for efficiently mapping of request classes to worker threads.
A comprehensive set of experiments showed that early scheduling improves the overall system performance. 

It is important to observe that the early scheduler does not provide the best degree of concurrency, but since decisions are simple and fast it outperforms other approaches that need a more complex processing 
to allow more parallelism in the execution. 
The existence (or not) of an optimization model that combines early scheduling and unrestricted concurrency is an open problem.




 
\bibliographystyle{abbrv}
\bibliography{IEEEabrv,referencias,AMPL}

\section*{A - Appendix - Correctness}
\label{proofs}

We argue that any mapping that follows the rules presented in \S~\ref{sec:mapping} generates correct executions.
We substantiate our claim with a case analysis, starting with classes without external conflicts.

\begin{enumerate}

\item \emph{Class $c$ has no internal and no external conflicts.}\label{case1} 
Then any request $r \in c$ is independent of any other requests and can be dispatched to the input queue of any thread (assigned to $c$).      
According to R.1, such a thread exists.    
According to line 6 of Algorithm \ref{alg_workers}, requests are dequeued in FIFO order.
Since the class is $Cnc$, lines 15--16 execute the request without further synchronizing with other threads.

\item \emph{Class $c$ has internal conflicts but no external conflicts.}\label{case2} 
Then, by rule R.2, request $r \in c$ is scheduled to be executed in all threads associated to $c$.
According to Al\-gorithm \ref{alg_scheduler}, requests are enqueued to threads in the order they are delivered.
According to Algorithm \ref{alg_workers}, lines 8--14, these threads synchronize to execute $r$, i.e., conflicting requests are executed respecting the delivery order.

\newcounter{enumTemp}
\setcounter{enumTemp}{\theenumi}
\end{enumerate}

Now we consider two conflicting classes. 

\begin{enumerate}
\setcounter{enumi}{\theenumTemp}

\item \emph{Class $c_1$ has no internal conflicts, but conflicts with $c_2$.}\label{case3} 
In such a case, by rule R.3, one of the classes must be sequential.   
Without loss of generality, assume $c_2$ is sequential and $c_1$ is concurrent.       
Since 
(i) the scheduler handles one request at a time (see Algorithm \ref{alg_scheduler}, line 3); 
(ii) it dispatches the request to all threads of $c_2$ (see lines 5--6); 
(iii) threads synchronize to execute (as already discussed in case \ref{case2}); and 
(iv) the threads that implement $c_1$ are contained in the set of threads that implement $c_2$ (from R.4), it follows that every request from $c_1$ is executed before or after any $c_2$'s request, according to the delivery order.
Due to the total order of atomic broadcast, all replicas impose the same order among $c_1$'s and $c_2$'s requests. 
Notice that this holds even though requests in $c_1$ are concurrent.

\item \emph{Classes $c_1$ and $c_2$ have internal conflicts and conflict with each other.}\label{case4} 
According to restriction R.5, these classes have at least one common thread $t_x$.
As in Algorithm \ref{alg_workers}, lines 5--14: (i) the input queue of $t_x$ has requests from both classes preserving the delivery order; 
(ii) according to R.2 and lines 7--14, $t_x$ will synchronize with all threads implementing $c_1$ to execute $c_1$'s requests and with all threads implementing $c_2$ to execute $c_2$'s requests.
This implies that $t_x$ imposes $c_1$'s and $c_2$'s requests to be executed in the replica sequentially, according to their delivery oder.  
Due to the total order of atomic broadcast, all replicas impose the same order of $c_1$'s and $c_2$'s requests. 

\newcounter{enumTempB}
\setcounter{enumTempB}{\theenumi} 
\end{enumerate}
 
Now we generalize to consider a system with an arbitrary request classes definition.
From Def. \ref{def.rClasses}, each request is associated to one class $c$ only.

Let $c$ be a concurrent class that conflicts with a number of other classes, the mapping imposes that all other classes 
are sequential (R.3) and include $c$'s threads (R.4).  As already discussed in case \ref{case3}, $c$'s requests are ordered w.r.t. requests from each of the other classes. 

Let $c$ be a sequential class that conflicts with a number of other classes.  The mapping states that each of these conflicting classes may be both sequential or concurrent (R.3).   
From cases \ref{case3} and \ref{case4}, requests from different classes, respectively concurrent and sequential, are processed according to the delivery order w.r.t. $c$'s requests.  

Since 
(i) the delivery order has no dependency cycles (i.e., a request $r_n$ can only depend on previously enqueued requests $r_m$, $m<n$); 
(ii) the request queue of each of the worker threads preserves this order; and
(iii) each thread processes its requests sequentially,
then request dependencies among worker threads is acyclic.
With this it is always possible to find a lowest element that does not depend on any other to be executed, and there is no deadlock.

We now show that any execution of early scheduling ensures linearizability.
From \S\ref{sec:model}, an execution is linearizable if there is a way to reorder the client requests in a sequence that (i) respects the semantics of the requests, as defined in their sequential specifications, and (ii) respects the real-time ordering of requests across all clients.  

(i.a)
Consider two independent requests $r_x$ and $r_y$. The execution of one does not affect the other and thus they can be executed in any relative order. 
As already discussed (cases~\ref{case1} and \ref{case2}), either $r_x$ and $r_y$ belong to a same concurrent class or they belong to different non-conflicting classes. In either case, these requests will be scheduled to execute without any synchronization between them.

(i.b)
Consider two dependent requests $r_x$ and $r_y$.  As already discussed, either $r_x$ and $r_y$ belong to a same internally conflicting class or they belong to different, conflicting classes.  
In any of those, there are common threads that synchronize to serialize the execution of the requests according to the order provided by the atomic broadcast. 
 Thus, the execution is sequential and respects the order provided by the atomic broadcast across replicas.
 Since their execution is sequential, their semantics is satisfied.   The order across replicas ensures consistent states.

(ii) Concerning the real-time constraints among $r_x$ and $r_y$, consider $r_x$ precedes $r_y$, i.e. $r_x$ finishes before $r_y$ starts.
Since before $r_x$ executes it must have been broadcast, this means that $r_x$ was delivered before $r_y$.   Therefore the delivery order satisfies the real-time constraints among the requests.

\end{document}